\documentclass[11pt ]{article}
\usepackage{amsmath}
\usepackage{graphicx}
\usepackage{hyperref}
\usepackage{lmodern}
\usepackage{amssymb}
\usepackage{booktabs}

\newcommand{\be}{\begin{equation}}
\newcommand{\ee}{\end{equation}}
\newcommand{\bea}{\begin{eqnarray}}
\newcommand{\eea}{\end{eqnarray}}

\renewcommand{\theequation}
{\arabic{section}.\arabic{equation}}

\makeatletter
\def\eqnarray{ \stepcounter{equation} \let\@currentlabel=\theequation
 \global\@eqnswtrue
 \global\@eqcnt\z@
 \tabskip\@centering
 \let\\=\@eqncr
 $$\halign to \displaywidth\bgroup\@eqnsel\hskip\@centering
 $\displaystyle\tabskip\z@{##}$&\global\@eqcnt\@ne
 \hfil$\displaystyle{{}##{}}$\hfil
 &\global\@eqcnt\tw@$\displaystyle\tabskip\z@{##}$\hfil
 \tabskip\@centering&\llap{##}\tabskip\z@\cr}
\makeatother

\makeatletter
\def\@arrayacol{\edef\@preamble{\@preamble \hskip .5\arraycolsep}}
\def\array{\let\@acol\@arrayacol \let\@classz\@arrayclassz
\let\@classiv\@arrayclassiv \let\\\@arraycr\def\@halignto{}\@tabarray}
\makeatother

\makeatletter
\newcounter{subeqncnt}
\def\thesubeqncnt{\alph{subeqncnt}}
\def\subequations{\begingroup%
   \stepcounter{equation}\edef\@tempa{\theequation}%
   \let\c@equation\c@subeqncnt\c@subeqncnt\z@
   \edef\theequation{\@tempa\noexpand\thesubeqncnt}}

\makeatother

\newcommand{\nn}{\nonumber}

\def\CC {{\cal C}}

\def\CM {{\cal M}}

\def\CZ {{\cal Z}}

\def\Det{\mathop {\rm Det}}

\begin{document}

\setlength{\baselineskip}{6mm}
\begin{titlepage}
\begin{flushright}

{\tt NRCPS-HE-05-2017} \\

\end{flushright}

\begin{center}
{\Large ~\\{\it

 Hyperbolic  Anosov  C-systems\\
 \vspace{0.5cm}
Exponential Decay of 
Correlation Functions\\

\vspace{0.5cm}
}

}

\vspace{1cm}

 {\sl  George Savvidy  and Konstantin Savvidy

 \bigskip
 \centerline{${}$ \sl Institute of Nuclear and Particle Physics}
\centerline{${}$ \sl Demokritos National Research Center, Ag. Paraskevi,  Athens, Greece}
\bigskip

}
\end{center}
\vspace{30pt}

\centerline{{\bf Abstract}}
The  uniformly hyperbolic Anosov  C-systems defined on a torus have 
exponential instability of their  trajectories,  and 
as such C-systems have mixing of all orders and
nonzero Kolmogorov entropy. 
The mixing property of all orders means that 
all its correlation functions  tend to zero and the 
question of a fundamental interest is a speed at which they 
tend to zero. It was proven that 
 the speed of decay in the C-systems is exponential, that is, 
the observables on the phase space become independent and uncorrelated exponentially fast. 
It is important to specify the properties of the C-system which quantify the exponential decay of correlations. We have found that 
the upper bound on the exponential decay of the correlation functions 
universally depends on the value of a system entropy.
A quintessence of the analyses is that local and homogeneous instability of the C-system
phase space trajectories translated into the exponential 
decay of the correlation functions
at the rate which is proportional to the  Kolmogorov entropy,  one of the fundamental characteristics of the Anosov automorphisms. This result allows to define the decorrrelation and relaxation times of  a C-system 
in terms of its entropy and 
characterise the statistical properties of a broad class of dynamical systems,  
including pseudorandom number generators and gravitational systems. 
\vspace{12pt}

\noindent

\end{titlepage}

\section{\it Introduction}
A  uniformly hyperbolic  Anosov  C-systems defined on a torus  
have exponential instability of all trajectories \cite{anosov}
and as such have {\it mixing of all orders and
nonzero Kolmogorov entropy} \cite{anosov,kolmo,kolmo1,sinai3,rokhlin,rokhlin2,sinai4,gines}.
The statistical properties of a deterministic 
dynamical system essentially depend 
on behaviour of the  correlation 
functions defined on a corresponding phase space.  The question of 
a fundamental interest is a speed at which 
 correlation functions of C-systems tend to zero \cite{bowen0,ruelle,yangmillsmech,yer1986a,Savvidy:1982jk,body}.  It was proven that for the 
hyperbolic Anosov C-systems the speed of decay is exponential, that is, 
the observables on the phase space become independent and uncorrelated exponentially fast \cite{Collet,moore,chernov1,young,simic,chernov,dolgopyat,chernov2,tsujiit3,tsujiit2,tsujiit1,tsujiit,faure,faure1}. 
{\it It is important to specify the properties of the C-system which quantify the exponential decay of the correlation functions}.

In this paper we shall study statistical properties of 
observables $\{f(x)\}$ defined on an $N$-dimensional torus phase space $\CM$ (  $ x \in \CM $) of the Anosov C-system diffeomorthisms  $T$ and specify the rate at which 
the exponential decay takes place. 
 The  statistical properties of the deterministic dynamical system defined by the map 
 $\{ \forall~ x \in \CM: x \rightarrow x_n=T^n x \}$ 
 are characterised by  the behaviour of the  corresponding correlation functions $D_n(f,g)$. 
The very fact that the C-systems have  mixing of all orders
means that the correlation function 
of any two observables $f(x)$ and $g(x)$ tends to zero\footnote{ In fact,  correlation functions of any number
of observables $\langle f_1(T^{n_1} x) f_2(T^{n_2} x) ....f_r(T^{n_r} x)\rangle$ of a C-system tend to zero as $n_{i} \rightarrow \infty$ \cite{anosov,Savvidy:1982jk}. We denote by 
$\langle f \rangle$ the average  value of a function $f$ with respect to 
invariant measure on $\CM$.} when iteration/interaction time $t=n$ tends to infinity $n \rightarrow \infty$:
\be
\label{correlations0}
D_n(f,g)= \langle f(x) g(T^n x) \rangle - \langle f(x)\rangle \langle g( x) \rangle
\rightarrow 0. \nn
\ee
This function measures the dependence between the values of 
$f(x)$ at zero time and values of $g(x)$ at the time $n$ and 
tells that the overlapping integral between observables $f(x)$ and $g(T^n x)$ tends to zero, so that they become independent and uncorrelated.  {\it A fundamental question which was raised in this respect is a question of a speed at which 
a correlation function $D_n(f,g)$ tends to zero in} (\ref{correlations0}). It was proven in the mathematical literature \cite{Collet,moore,chernov1,young,simic,chernov,dolgopyat,chernov2,tsujiit3,tsujiit2,tsujiit1,tsujiit,faure,faure1} that 
for the hyperbolic Anosov diffeomorphisms and continuous flows the speed of decay is exponential:
\be\label{exponen7}
\vert D_n(f,g) \vert = \vert \langle f(x) g(T^n x) \rangle - \langle f(x)\rangle \langle g( x) \rangle \vert ~~\leq~~ C(f,g)~ \Lambda^n = C(f,g)~ \exp{(- n~ \log1/\Lambda)},  \nn
\ee
where the exponential upper bound
on the correlation functions  $\Lambda_{ (T,f,g) } < 1   $ depends on properties of the dynamical system $T$ and of the observables $f(x)$ and $g(x)$.  
A constant $C(f,g)$
depends  only on the observables $f(x)$ and $g(x)$. This outstanding result tells that the observables on the phase space of a C-system become uncorrelated exponentially fast and represent 
independent random variables. In statistical physics 
the autocorrelation functions $D_n(f,f)$  define the important 
physical properties of a dynamical  system  $T$,
such as its relaxation time $\tau$, as well as temperature,  diffusion, viscosity and other macroscopic characteristics  \cite{krilov,yer1986a,body,Gibbons:2015qja}.

 It is important  to express 
 $\Lambda$ and  the relaxation time $\tau$ 
 in terms of C-system quantitative characteristics. For that one should specify the properties of a C-system which quantify the exponential decay of the correlation functions in (\ref{exponen7}).  We have found that 
the upper bound on the exponential decay of the correlation functions 
is universal and is defined  by the value of the system entropy $h(T)$:  
\be\label{main}
\vert D_n(f,g) \vert \leq  C~  e^{-n h(T) \nu },
\ee
where $C(f,g)$ and $\nu(f,g)$ depend only on observables and are 
positive numbers. This result  allows to define the decorrelation time 
$\tau_0$ for a physical observable $f(x)$ as  
\be
\tau_0 = {1\over h(T) \nu_f}~~.
\ee
{\it A local and homogeneous  instability of the C-system
phase trajectories is translated into the exponential decay of the correlation functions
at rate which is expressed in terms of the system entropy $h(T)$.}
The expression of the decorrelation  time in terms of system entropy allows to characterise statistical properties of a broad class of dynamical systems,   including gravitational systems and pseudorandom number generators  
\cite{yer1986a,body,konstantin,Savvidy:2015ida,Savvidy:2015jva}. 

When the dimension $N$ of the C-system (\ref{eq:matrix}) on a torus is increasing, 
its index $\nu_f$
is increasing linearly with dimension $\nu_f = 2p N$, where $p$ is the order 
of smoothness of the observable/function $f(x)$. The entropy $h(T)$ 
of the C-system $(\ref{linear})$ increases linearly as well $h(T)= {2 \over \pi} N$, therefore 
\be
\tau_0 = {\pi \over 4 p N^2}~~.
\ee
Considering a set of initial trajectories occupying a small volume $\delta v_0$
in the phase space of a C-system, one can  ask how fast this
small phase volume will be  uniformly distributed over the whole phase space.  
This  characteristic 
time interval $\tau$ defines the relaxation time at which the system reaches a stationary distribution. 
Because the entropy defines the expansion rate of the phase space volume 
one can  derive that \cite{yer1986a}
\be
\tau = {1\over h(T)} \ln {1\over \delta v_0}.
\ee  
Thus there are three characteristic time scales associated with the C-system
\cite{yer1986a}:
 \be
  \begin{pmatrix}
 Decorrelation~ time~\\
 \tau_0 ={\pi \over 4 p N^2}
\end{pmatrix}  < 
\begin{pmatrix}
 Interaction~ time~  \\
 t_{int} =n=1
\end{pmatrix} <
\begin{pmatrix}
 Stationary~ distribution~ time  \\
\tau={1\over h(T)} \ln {1\over \delta v_0}
\end{pmatrix}.
 \ee
This result defines important physical characteristics of the C-systems 
and measures the "level of chaos" developed in the system and justifies 
the statistical/probabilistic description of the system \cite{krilov}.
Indeed, the appearance of well developed statistical properties 
has important consequences in the form of the central limit
theorem for Anosov diffeomorphysms.   The time average  of the observable  $f(x)$ on $\CM$ 
 \be\label{timeav}
\bar{f}_n(x) = {1 \over n} \sum^{n-1}_{i=0} f(T^i x) \nn
\ee
behaves  as a superposition of quantities which are statistically independent \cite{leonov}.  It has been proven  that
the fluctuations of the time averages (\ref{timeav}) from the phase space average 
\be\label{spaceav}
\langle f \rangle  = \int_{\CM} f(x) d \mu(x)  \nn
 \ee
multiplied by $\sqrt{n}$ 
have at large $n \rightarrow \infty$ the Gaussian distribution \cite{mac,leonov,chernov3,rozanov,rather}:
 \be\label{gauss}
\lim_{n \rightarrow \infty}\mu \bigg\{ x	:\sqrt{n}   \Bigg( \bar{f}_n(x)  - \langle f \rangle  \Bigg) < z    \bigg\}
= {1 \over \sqrt{2 \pi \sigma^2_f}}\int^{z}_{-\infty} e^{-{y^2 \over 2 \sigma^2_f}} dy~,\nn
\ee
where the value of the standard deviation  $\sigma_f$ is a sum
\be
\sigma^2_f = \langle f^2(x)  \rangle-
\langle f(x)  \rangle^2  +  2 \sum^{+ \infty}_{n=1} [\langle f(T^n x) f(x)  \rangle-
\langle f(x)  \rangle^2 ]. \nn
\ee
Using our result (\ref{main})  one can 
explicitly  estimate the standard deviation in terms of entropy:
\be
\sigma^2_f \leq  
C~ {1+e^{ -h(T) \nu} \over 1- e^{  -h(T) \nu}  }. \nn
\ee 

The earlier publications concerning the application of the modern results of the ergodic theory to concrete physical systems can be found in \cite{krilov,yangmillsmech,yer1986a,Savvidy:1982jk,body,Gibbons:2015qja}. These articles contain review material as well. 
The present paper is organised as follows. In section two we shall overview the basic properties of a C-system defined on a high dimensional  torus, its spectral properties and its entropy.  In section three we shall calculate the correlation functions 
and shall express the upper bound on the correlation functions in terms of entropy
for the system on two-dimensional torus.  In the  fourth section we shall 
extend these results to the high dimensional C-systems and shall define 
three characteristic time scales associated with the C-systems. In section 
five the time 
scales associated with the MIXMAX pseudorandom number generators 
\cite{yer1986a,konstantin,Savvidy:2015ida, Savvidy:2015jva}
will  be estimated. In conclusion we summaries the results.  
  
\section{\it C-sytems on a Torus}

A particular system chosen for investigation is the one realising linear automorphisms of the unit hypercube $\CM^N$ in Euclidean space $E^N$ with coordinates $ (x_1,...,x_N)$\cite{anosov,yer1986a,konstantin,Savvidy:2015ida}:
\be
\label{eq:rec}
x_i^{(k+1)} = \sum_{j=1}^N T_{ij} \, x_j^{(k)} ~~~~~\textrm{mod}~ 1,~~~~~~~~~k=0,1,2,...
\ee
where the components of the vector $x^{(k)}$ are defined as 
$
x^{(k)}= (x^{(k)}_1,...,x^{(k)}_N).
$
The phase space $\CM^N$ of the systems   (\ref{eq:rec} ) can also be 
considered a $N$-dimensional torus \cite{anosov,yer1986a,konstantin,Savvidy:2015ida}, 
appearing at  factorisation of the 
Euclidean space $E^N$ with coordinates $x= (x_1,...,x_N)$ 
over an integer lattice $\CZ^N$. The operator $T$ acts on the 
initial vector $x^{(0)}$ and produces a phase space trajectory $x^{(n)}=  T^n x^{(0)}$
on a torus. 

\begin{figure}
\begin{center}
\includegraphics[width=5cm]{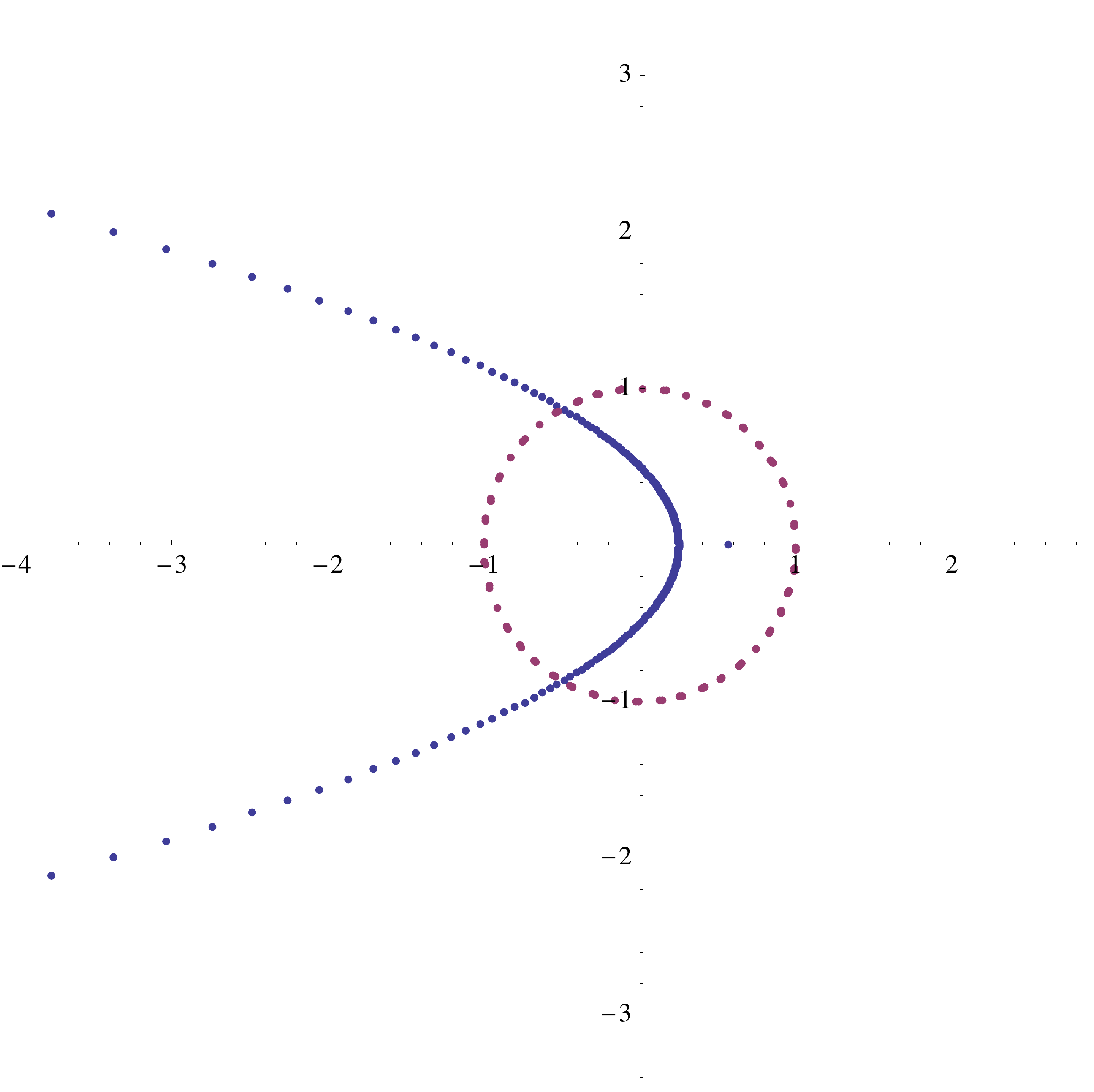}~~~~~~~~~~
\includegraphics[width=5cm]{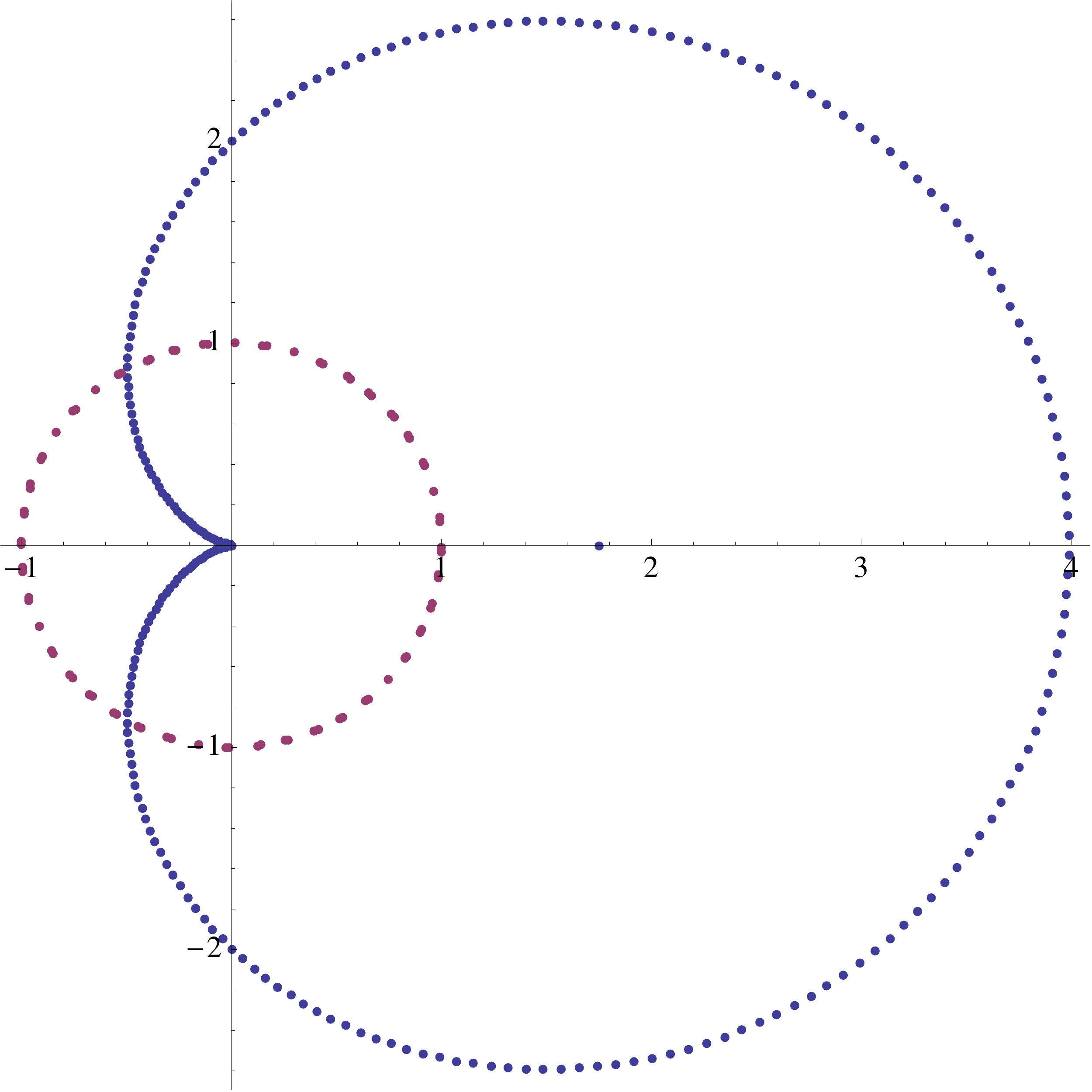}
\caption{
On the left figure is the distribution of the eigenvalues of the operator $T$
and on the right is the  distribution of the eigenvalue of the inverse 
operator $T^{-1}$ for the $N=256$.
}
\label{fig1}
\end{center}
\end{figure}

The dynamical system defined by the integer matrix $T$
has a determinant equal to one $\Det T =1$ and has no eigenvalues on the unit circle
\cite{anosov}. The  spectrum $\{ {\lambda_1},...,
\lambda_N \}$ of the matrix $T$ fulfils therefore the following
two conditions:
\bea\label{mmatrix}
1)~\Det  T=  {\lambda_1}\,{\lambda_2}...{\lambda_N}=1,~~~~~
2)~~\vert {\lambda_i} \vert \neq 1, ~~~\forall ~~i .~~~~~~
\eea
The Liouville's measure $d\mu = dx_1...dx_N$ is invariant under the action of $T$,
and  $T$ is an automorphism of  the unit hypercube  onto itself.
The conditions (\ref{mmatrix}) on the eigenvalues of the matrix $T$ are  sufficient
to prove that the system represents  an Anosov C-system \cite{anosov} and therefore as such it also represents a  Kolmogorov K-system \cite{kolmo,kolmo1,sinai3,rokhlin,rokhlin2}
with mixing  of  all orders and of nonzero entropy.
The eigenvalues of the matrix $T$ can be divided
into two sets   $\{ \lambda_{\alpha}  \} $ and $\{  \lambda_{\beta }  \} $
with modulus smaller and larger than one:
\bea\label{eigenvalues}
0 <  \vert \lambda_{\alpha} \vert   < 1   & \textrm{ for } \alpha=1...d\nn\\
1 <  \vert \lambda_{\beta} \vert  < \infty & \textrm{ for } \beta=d{+}1...N . 
\eea
There exist two  hyperplanes $ X= \{X_{\alpha} \}$ and $ Y= \{Y_{\beta} \}$
which are  spanned by the
corresponding eigenvectors  $\{ e_{\alpha}  \}$ and  $\{ e_{\beta}  \}$.
These invariant planes define invariant spaces on which the phase trajectories 
are  expanding and contracting
under the transformation $T$ at an exponential rate.  
The C-system (\ref{eq:rec})  has a nonzero Kolmogorov entropy $h(T)$  \cite{anosov,sinai3,rokhlin2,sinai4,gines,Savvidy:2015ida}:
\be\label{entropyofA}
h(A) = \sum_{\beta   } \ln \vert \lambda_{\beta} \vert
\ee
which is expressed in terms of the eigenvalues $\lambda_{\beta}$ of the operator $T$ .
{\it The entropy quantitatively characterises the instability of a 
C-system trajectories and  its value depends on the
spectral properties of the evolution operator $T$}.  
We shall consider a family of operators of dimension  $N$
introduced in \cite{konstantin}:
\be
\label{eq:matrix}
T =
   \begin{pmatrix}
      1 & 1 & 1 & 1 & ... &1& 1 \\
      1 & 2 & 1 & 1 & ... &1& 1 \\
            1 & 2 & 2 & 1 & ... &1& 1 \\
      1 & 4 & 3 & 2 &   ... &1& 1 \\
      &&&...&&&\\
      1 & N & N{-}1 &  ~N{-}2 & ... & 3 & 2
   \end{pmatrix}
\ee
The operator $T$  fulfils the C-condition (\ref{mmatrix}) and represents a C-system \cite{yer1986a}. The spectrum of the operator $T$   and of its inverse $T^{-1}$
are presented in Fig.\ref{fig1} \cite{konstantin,Savvidy:2015jva}.  
The entropy of the C-system $T$ can be calculated for
large values of $N$  \cite{konstantin,Savvidy:2015jva}:
\be\label{linear}
h(A)= \sum_{\beta   } \ln \vert { \lambda_{\beta} }\vert \approx
{2\over \pi}~ N
\ee
and increases linearly with the dimension $N$.
Our aim is to study  the behaviour of the observables $\{f(x)\}$ defined 
on the torus phase space  $\CM^N$ of the dynamical system $T$  (\ref{eq:matrix})
and, in particular, a speed at which the correlation functions decay.  

\section{\it Correlation Functions}

The general form of the correlations we are intending to consider are: 
\be
\label{correlations}
D_n(f,g)= \langle f(x) g(T^n x) \rangle - \langle f(x)\rangle \langle g(x) \rangle,
\ee
where the $f(x)$ and $g(x)$ are the observables/functions defined on the torus 
phase space $\CM^N$. We shall consider the observables belonging 
to a general class of functions which are  p-times differentiable $f,g \in C^p$, where p is an integer,  or functions which are in the  $\alpha$-H\"older class.  

\begin{figure}
\begin{center}
\includegraphics[width=8.7cm]{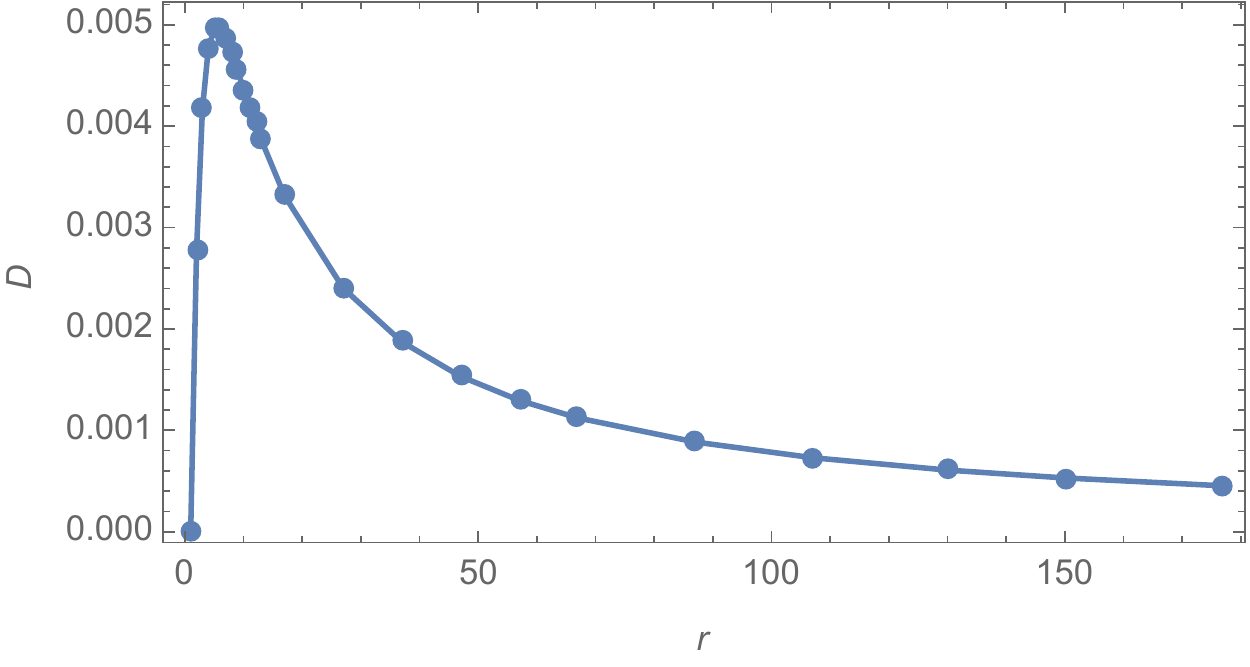}~~~
\includegraphics[width=7cm]{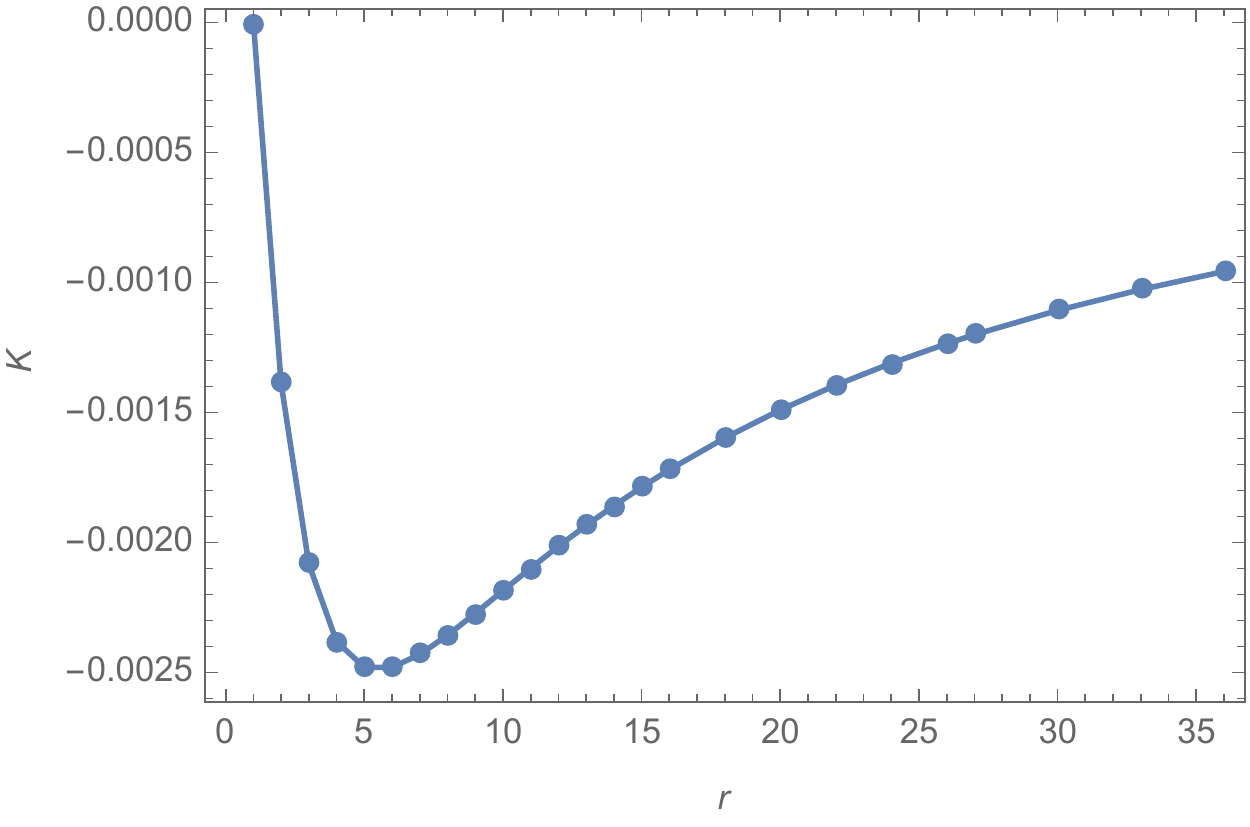}~~~~~~~
\caption{The correlation function $D_1(r)=<x_1 x_2^r \{x_1+x_2\}>-<x_1 x_2^r>< \{x_1+x_2\}>$ and 
the correlator    
$K_1(r)=<x_1 x_2  \{x_1+2x_2\}^r>-<x_1 x_2>< \{x_1+2x_2\}^r>$, where 
$  \{x\}~  \equiv  x ~mod~1$.}
\label{fig3}  
\end{center}
\end{figure}

In this section the operator $T$ is a two-dimensional matrix , $N=2$  in (\ref{eq:matrix}):
\be
\label{eqmatrix3}
Tx =  
   \begin{pmatrix}  
      1 & 1   \\
      1 & 2
       \end{pmatrix} \begin{pmatrix}  
     x_1   \\
      x_2
       \end{pmatrix}~~mod~1~,~~~~~~~
       \begin{pmatrix}  
     x_1   \\
      x_2
       \end{pmatrix} 
        \rightarrow  
         \begin{pmatrix}  
   \{ x_1 + x_2\}  \\
      \{ x_1 + 2 x_2\}
       \end{pmatrix} ,
     \ee
where $  \{x\}~  \equiv  x ~mod~1$.
To have an idea about how the correlation $D_n$ behaves as a function of the
time step $n$
we shall calculate  a "one-step"  correlation  $D_1(f,g)$ when the 
observables are separated by one unit of time $n=1$ and  $f $ and $g$
are some polynomial functions. A simple  example will be of the form 
\be\label{onestep}
D_1(f,g)= <f(x)~ Tx > - <f(x)><x>\nn,
\ee
where  $f= x_1 x_2^r$ or  $f=(x_1 x_2)^r$, $r=0,1,...$ and $g=x$. The result
of calculations is presented on 
Fig.\ref{fig3}  and demonstrates that a one-step 
correlation decreases, $D_1(r) \rightarrow 0$,  as the order of the polynomials $f$
increases, $r \rightarrow \infty$. 
In order to confirm this behaviour analytically let us consider the 
observables of the form 
\bea
\label{observe1}
f(x) =  \sum^{\infty}_{i_1,i_2=1} a_{i_1 i_2} \sin(2 \pi i_1  x_1  )
\cos(2 \pi i_2   x_2),~~~<f(x)>=0,~~~g(x)=x,~~~<g(x)>={1\over 2},\nn
\eea
where the numbers $(i_1,i_2)$ define the oscillation frequencies of the 
observable $f$.
The correlation function (\ref{onestep})  will take the form
\bea\label{fur1}
D_1(f,g)  =\sum^{\infty}_{i_1,i_2=1}a_{i_1 i_2} \int^1_0 dx_1 dx_2
\sin(2 \pi i_1  x_1  )
\cos(2 \pi i_2   x_2)  \{ x_1 + x_2\} =
-\sum^{\infty}_{r=1} {a_{rr}\over 4\pi r},~~~~
\eea
where we used a trigonometric representation of the $mod~1$ operation:
\be
(x_1 +x_2)~mod~1 \equiv \{x_1 +x_2\} =  {1\over 2} - {1\over \pi}\sum^{\infty}_{r=1}{1\over r}\sin(2 \pi r (x_1 +x_2)). \nn
\ee 
In order to estimate the 
behaviour of the Fourier coefficients in (\ref{fur1}) we shall consider the functions $f(x)$ which have $(2p_1,2p_2)$ continuous 
partial derivatives $f(x_1,x_2) \in \CC^{2p_1,2p_2}(\CM)$. The 
behaviour of the Fourier coefficients can be found performing a
 partial integrations and using the  periodicity of the functions $f(x +1) =f(x)$.  Thus  
 the Fourier coefficients can be represented in the form  \cite{Poincare1}
 \footnote{If a function is finite and continuous together with its $p-2$ 
 derivatives and if its $p-1$ derivative is bound and has finite 
 number of discontinuities, then there exists a finite constant $M_p $ such that $\vert a_i \vert 
 < {M_p/i^p}$ \cite{Poincare1}
.}:
\be
\label{poincare}
a_{i_1 i_2} = {4 (-1)^{p_1+p_2}\over (2\pi i_1)^{2p_1}(2\pi i_2)^{2p_2}}
\int^1_0 dx_1 dx_2  \sin(2 \pi i_1  x_1  )
\cos(2 \pi i_2   x_2)~ \partial^{(2p_1)}_{x_1} \partial^{(2p_2)}_{x_2} f(x_1,x_2).
\ee
This representation allows to estimate the one-step  correlation function $D_1(f,g)$:
\bea
\vert D_1(f,g)\vert = \vert <f(x)~ T x> - <f(x)><x> \vert = \vert \sum^{\infty}_{r=1} {a_{rr}\over 4\pi r}\vert   \leq
\sum^{\infty}_{r=1}    { 2 M_p  \over   (2\pi r)^{2p_1+2p_2+1}} ~,
\eea
where 
\be
\vert \partial^{(2p_1)}_{x_1} \partial^{(2p_2)}_{x_2} f(x_1,x_2) \vert \leq M_p . \nn
\ee
This result confirms our numerical observation demonstrated on  
Fig.\ref{fig3} that 
the one-step correlation $D_1(f,g)$ decreases when the oscillation frequency 
  $r$ of the observable $f$ increases:
\be
D_1(r) \sim {1\over r^{2p_1+2p_2+1}} \rightarrow 0.
\ee
The  oscillations  frequencies of the observable $f(x)$ are defined by $(i_1,i_2) $ in (\ref{observe1}) and we are considering the limit when  $(i_1,i_2) \sim r \rightarrow \infty$.  
Performing  a similar calculation 
one can get convinced that the result holds  
for more general  functions $f(x)$ in (\ref{observe1}):
\be
\label{observe2}
f(x) =  \sum^{\infty}_{i_1,i_2=0}  a_{i_1 i_2} \cos(2 \pi i_1  x_1  )
\cos(2 \pi i_2   x_2) + \sum^{\infty}_{i_1,i_2=1}  b_{i_1 i_2} \cos(2 \pi i_1  x_1  )
\sin(2 \pi i_2   x_2)+...\nn
\ee
The other independent correlation function of $Tx$  is   of the form
 \bea
<f(x)~ T x>  =\sum^{\infty}_{i_1,i_2=1}  a_{i_1 i_2} \int^1_0 dx_1 dx_2
\sin(2 \pi i_1  x_1  )
\cos(2 \pi i_2   x_2)  \{ x_1 + 2 x_2\} =
-\sum^{\infty}_{r=1} {a_{r,r+2}\over 4\pi r}.~~~~\nn
\eea
As one can see, the second index of the Fourier coefficient 
 was shifted by two units $a_{r,r+2}$.  Examining its behaviour by
using (\ref{poincare}) we conclude that the coefficients decay faster: 
\be
\vert D_1(f,g)\vert =   \vert \sum^{\infty}_{r=1}
 {a_{r,r+2}\over 4\pi r}\vert   \leq
\sum^{\infty}_{r=1}    { 2M_p  \over   (2\pi r)^{2p_1+1}(2\pi (r+2))^{2p_2}}.
\ee

Of our primary interest is to find out the behaviour of the correlations $D_n(f,g)$ for  
observables $f(x)$ and $g(T^nx)$  separated by  $n$ time steps (\ref{correlations}). 
For that one 
should generalise our previous calculation in two directions, considering 
 $n \geq 1$ and the general  functions $g(x)$ in (\ref{correlations}).
Thus we shall consider a Fourier representation  of the function $g(x)$ on a torus of the form
\be
g(x_1,x_2)= \sum^{\infty}_{j_1,j_2=1}  b_{j_1j_2}\cos(2 \pi j_1 x_1)\cos(2 \pi j_2 x_2).
\ee
In order to calculate the observable $g(x)$ after $n$-steps  $g(T^nx)=g(\{x_1+x_2\},\{x_1+2x_2\})$ we have to define a $mod~1$ operation acting on a nonlinear  
function $g(x)$. The $mod~1$ operation can be easily realised on trigonometric functions, since ~~
$
\sin(2 \pi j x)=\sin(2 \pi j (\{x\} + integer ))= \sin(2 \pi j \{x\}),$
$\cos(2 \pi j x)=\cos(2 \pi j (\{x\} + integer ))= \cos(2 \pi j \{x\})$,
therefore 
\be
g(\{x_1\},\{x_2\})= \sum^{\infty}_{j_1,j_2=1}  b_{j_1 j_2}\cos(2 \pi j_1 x_1)\cos(2 \pi j_2 x_2).
\ee
This is an important observation for a successful calculation of $D_n(f,g)$, because if one considers a polynomial expansion of $f$ and $g$, then the $mod~1$ operation on polynomials is much more difficult to execute. Thus we shall consider a Fourier representation  of the observables:
\bea
&&f(x) =  \sum^{\infty}_{i_1,i_2=1}  a_{i_1 i_2} \cos(2 \pi i_1  x_1  )
\cos(2 \pi i_2   x_2), ~~~~
\{T^nx\} =  
   \begin{pmatrix}  
      \{a_n x_1+b_n x_2  \} \\
      \{c_nx_1+d_n x_2\}
       \end{pmatrix},
\\
&& g(T^n x) 
=\sum^{\infty}_{j_1,j_2=1} b_{j_1 j_2}
\cos(2 \pi j_1 (a_n x_1 +b_n x_2) )\cos(2 \pi j_2 (c_n x_1 +d_n x_2)),\nn
\eea
where the coefficients $(a_n, b_n, c_n,  d_n)$ define the n'th power 
of the operator $T$ and can be expressed in terms of  Fibonacci numbers $F_{2n}$:
\be
\label{eqmatrix2}
T^n =  
   \begin{pmatrix}  
      a_n & b_n   \\
      c_n & d_n
       \end{pmatrix}  =
\begin{pmatrix}  
      F_{2n} -F_{2n-2}& F_{2n}   \\
      F_{2n} & 2F_{2n}-F_{2n-2}
       \end{pmatrix}  ,~~~~~~F_{2n}= {\lambda^n - \lambda^{-n}\over \sqrt{5}}    ~~.  
\ee
The $\lambda = {3+\sqrt{5}\over 2}>1 $ is the eigenvalue of the matrix $T$ and 
\be\label{deter}
det T^n = a_n d_n -b_n c_n=1,~~~~d_n > b_n=c_n > a_n.
\ee
Using the above formulas we can express the general correlation functions in the form
\bea
\label{cor1}
&&~~~~~~~~~~~~~~~~~~~~~~~~~~~~~~~~~~ \langle f(x) g(T^n x) \rangle 
 =  \\
&&= \sum^{\infty}_{i_1,i_2,j_1,j_2=1} a_{i_1 i_2}b_{j_1 j_2}
 {1\over 8}\bigg(\delta_{i_1,j_1 a_n + j_2 c_n} \delta_{i_2, j_1 b_n + j_2 d_n } + 
 \delta_{j_1 a_n, i_1 + j_2 c_n}   \delta_{j_1 b_n, i_2 + j_2 d_n} +
\nn\\
&&~~~~~~~~~~~~~~~~~~~~~~ +\delta_{j_1 a_n, i_1 + j_2 c_n}   \delta_{j_2 d_n, i_2 + j_1 b_n} 
+  \delta_{j_2 c_n, i_1 + j_1 a_n}   \delta_{j_2 d_n, i_2 + j_1 b_n} +
 \delta_{j_2 c_n, i_1 + j_1 a_n}   \delta_{j_1 b_n, i_2 + j_2 d_n} \bigg) \nn .
 \eea
 In order to execute the last four delta functions in (\ref{cor1}) one should solve the linear equations 
 \be
 \begin{array}{ll}
a_n j_1 - c_n j_2 =i_1\\
\pm b_n j_1 \mp  d_n j_2 =i_2
\end{array},~~~~~
\begin{array}{ll}
- a_n j_1 + c_n j_2 =i_1\\
\mp b_n j_1 \pm  d_n j_2 =i_2
\end{array}
 \ee
with respect  to the $j_1,j_2$. All these equations have unique solutions because the corresponding determinants 
are equal to $\pm 1$  (\ref{deter}) and we have to select the 
solutions  with positive $j_1,j_2$. The last delta function in (\ref{cor1}) has no 
positive solutions and therefore does not contribute to the correlation.
Thus we have  
 \bea\label{correlator2}
 && \langle f(x) g(T^n x) \rangle  ={1\over 8}\sum^{\infty}_{j_1,j_2=1} 
a_{j_1 a_n + j_2 c_n,  j_1 b_n + j_2 d_n} b_{j_1 j_2}  +  
{1\over 8}\sum^{\infty}_{i_1,i_2=1} 
a_{i_1,i_2}  b_{ i_1 d_n + i_2 c_n, i_1b_n+i_2 a_n} +\\
&&+{1\over 8}\sum^{\infty}_{i_1 d_n >i_2 c_n;i_1b_n >i_2 a_n} 
a_{i_1,i_2} b_{ i_1 d_n - i_2 c_n, i_1b_n-i_2 a_n}+  {1\over 8}\sum_{ i_2 c_n>i_1 d_n;i_2 a_n >i_1b_n} 
a_{i_1,i_2}
  b_{ -i_1 d_n + i_2 c_n, -i_1b_n+i_2 a_n}\nn.
 \eea
The subtraction terms in (\ref{correlations}) are:
\bea\label{cor2}
&&  <f(x)> =\sum_{i_1,i_2 =1} a_{i_1 i_2} \int^1_0 dx_1 dx_2
\cos(2 \pi i_1  x_1  )
\cos(2 \pi i_2   x_2)   = 0, \nn\\
&&<g(x)>=\sum_{j_1,j_2=1} b_{j_1 j_2}\int^1_0 dx_1 dx_2
 \cos(2 \pi j_1   x_1  )\cos(2 \pi j_2  x_1 )
=0 \nn,
\eea
and the total expression for the correlation will take the following form: 
\bea\label{realisation}
&&D_n(f,g)= <f(x) g(T^n x)> - <f(x)><g( x)> = \\
&& ={1\over 8}\sum^{\infty}_{j_1,j_2=1} 
a_{j_1 a_n + j_2 c_n,  j_1 b_n + j_2 d_n} b_{j_1 j_2}  +  
{1\over 8}\sum^{\infty}_{i_1,i_2=1} 
a_{i_1,i_2}  b_{ i_1 d_n + i_2 c_n, i_1b_n+i_2 a_n} +\nn\\
&&+{1\over 8}\sum^{\infty}_{i_1 d_n >i_2 c_n ; ~i_1b_n >i_2 a_n } 
a_{i_1,i_2} b_{ i_1 d_n - i_2 c_n, ~i_1b_n-i_2 a_n}+  {1\over 8}\sum^{\infty}_{ i_2 c_n>i_1 d_n; ~i_2 a_n >i_1b_n} 
a_{i_1,i_2}
  b_{ -i_1 d_n + i_2 c_n, -i_1b_n+i_2 a_n}\nn .
\eea
We have to estimate each term in the above expression. 
Using the representation (\ref{poincare}) we shall evaluate  the first term 
in the correlator  $D_n(f,g)$:  
\bea
&&{1\over 8}\vert  \sum^{\infty}_{r_1,r_2=1}  a_{r_1 a_n + r_2 c_n,  r_1 b_n + r_2 d_n} b_{r_1 r_2}\vert   \leq \nn\\
&&2 \sum^{\infty}_{r_1,r_2=1}   {  M_p  \over   (2\pi (r_1 a_n + r_2 c_n)^{2p_1}(2\pi (r_1 b_n + r_2 d_n))^{2p_2}}  {  M_q \over   (2\pi r_1)^{2q_1}(2\pi r_2)^{2q_2}}  =\nn\\
&&{2\over (b_n)^{2p_1}(c_n)^{2p_2}}\sum^{\infty}_{r_1,r_2=1}  {  M_p  \over   (2\pi (r_2 +r_1 {a_n \over c_n} )^{2p_1}(2\pi (r_1  + r_2 {d_n \over b_n}))^{2p_2}}  {  M_q \over   (2\pi r_1)^{2q_1}(2\pi r_2)^{2q_2}}    \leq \nn\\
&& e^{-2p_1 \ln b_n - 2p_2 \ln c_n}~ \sum^{\infty}_{r_1,r_2=1}   {  M_p  \over   (2\pi r_2  )^{2p_1}(2\pi r_1 )^{2p_2}}  { 2  M_q \over   (2\pi r_1)^{2q_1}(2\pi r_2)^{2q_2}}   ~~,\nn
\eea
where 
\be
\vert \partial^{(2p_1)}_{x_1} \partial^{(2p_2)}_{x_2} f(x_1,x_2) \vert \leq M_p, ~~~~~~~
\vert \partial^{(2q_1)}_{x_1} \partial^{(2q_2)}_{x_2} g(x_1,x_2) \vert \leq M_q ~.
\ee
The logarithm in the exponent can be expressed as
$$
\ln b_n =\ln c_n= \ln F_{2n} =\ln {\lambda^n -\lambda^{-n} \over\sqrt{5}} = n \ln \lambda +
\ln(1-\lambda^{-2n}) - {1\over 2}\ln 5 ~> ~n \ln \lambda .
$$
Thus we  have 
\bea
{1\over 8}\vert  \sum^{\infty}_{r_1,r_2=1}  a_{r_1 a_n + r_2 c_n,  r_1 b_n + r_2 d_n} b_{r_1 r_2}\vert   \leq e^{-n (2p_1 +2p_2) \ln \lambda } ~\sum^{\infty}_{r_1,r_2=1}    { 2  M_p M_q \over   (2\pi r_1)^{2q_1+2p_2}(2\pi r_2)^{2p_1+2q_2}}   \nn
\eea
and remembering that the entropy of the system is 
$h(T)=  \ln \lambda$ (\ref{entropyofA}), we have for the first term 
\be
 {1\over 8} \vert  \sum^{\infty}_{r_1,r_2=1}  a_{r_1 a_n + r_2 c_n,  r_1 b_n + r_2 d_n} b_{r_1 r_2}\vert ~~~\leq  ~~~~ C_1(f,g)~ 
 e^{-n h(T) \nu_1(f,g)} ~~,
\ee
where the  numerical factors 
\bea
C_1(f,g) =\sum^{\infty}_{r_1,r_2=1}    { 2  M_p M_q \over   (2\pi r_1)^{2q_1+2p_2}(2\pi r_2)^{2p_1+2q_2}},~~~\nu_1(f,g) =2p_1 +2p_2  \nn
\eea
depend only on the observables  
and are independent of the system dynamics $T$.
 For the second term we have a similar estimate:
\bea
&& {1\over 8} \vert  \sum^{\infty}_{r_1,r_2=1} 
a_{r_1,r_2}  b_{ r_1 d_n + r_2 c_n, r_1b_n+r_2 a_n}\vert    \leq  \nn\\
&& \sum^{\infty}_{r_1,r_2=1}   
{ 2 M_p  \over   (2\pi r_1 )^{2p_1}(2\pi  r_2 )^{2p_2}} 
 {   M_q \over   (2\pi (  r_1 d_n + r_2 c_n   )   )^{2q_1}(2\pi (   r_1b_n+r_2 a_n   ))^{2q_2}}  =\nn\\
&&{2\over (b_n)^{2q_2}(c_n)^{2q_1}}\sum^{\infty}_{r_1,r_2=1}   
 {  M_p \over   (2\pi r_1)^{2p_1}(2\pi r_2)^{2p_2}}  
  {  M_q  \over   (2\pi (r_2 +r_1 {d_n \over c_n} )^{2q_1}(2\pi (r_1  + r_2 {a_n \over b_n}))^{2q_2}} \leq \nn\\
&& e^{-2q_2 \ln b_n - 2q_1 \ln c_n} \sum^{\infty}_{r_1,r_2=1}    { 2 M_p \over   (2\pi r_1)^{2p_1}(2\pi r_2)^{2p_2}}   {  M_q  \over   (2\pi r_2  )^{2q_1}(2\pi r_1 )^{2q_2}} 
\leq 
\nn\\ 
&&e^{-n (2q_1 +2q_2) \ln \lambda } ~\sum^{\infty}_{r_1,r_2=1}    { 2  M_p M_q \over   (2\pi r_1)^{2p_1+2q_2}(2\pi r_2)^{ 2q_1+2p_2}} ,  \nn
\eea
thus 
\be
 {1\over 8} \vert  \sum^{\infty}_{r_1,r_2=1} 
a_{r_1,r_2}  b_{ r_1 d_n + r_2 c_n, r_1b_n+r_2 a_n}\vert ~~~   \leq ~~~~ C_2(f,g)~   e^{-n  h(T) \nu_2(f,g)} ,
\ee
where the  numerical factors are: 
\bea
C_2(f,g) =\sum^{\infty}_{r_1,r_2=1}     { 2  M_p M_q \over   (2\pi r_1)^{2p_1+2q_2}(2\pi r_2)^{ 2q_1+2p_2}},~~~\nu_2(f,g)=  2q_1 +2q_2,   \nn
\eea 
and they are independent  of the system dynamics $T$.
The third and the fourth terms of $D_n(f,g)$ in (\ref{realisation}) are 
\bea
&&+{1\over 8}\sum^{\infty}_{i_1 d_n >i_2 c_n; i_1b_n >i_2 a_n} 
a_{i_1,i_2} b_{ i_1 d_n - i_2 c_n, i_1b_n-i_2 a_n}+  {1\over 8}\sum^{\infty}_{ i_2 c_n>i_1 d_n; i_2 a_n >i_1b_n } 
a_{i_1,i_2}
  b_{ -i_1 d_n + i_2 c_n, -i_1b_n+i_2 a_n}\nn
\eea
and using the representation (\ref{poincare}) for the Fourier coefficients one can find that 
\bea\label{third}
&& {1\over 8}\vert  \sum^{\infty}_{r_1 d_n >r_2 c_n; r_1b_n >r_2 a_n} 
a_{r_1,r_2} b_{ r_1 d_n - r_2 c_n, r_1b_n-r_2 a_n} \vert  \leq\nn\\
&&\leq 
\sum^{\infty}_{r_1 d_n >r_2 c_n; r_1b_n >r_2 a_n}   
 {  2 M_p \over   (2\pi r_1)^{2p_1}(2\pi r_2)^{2p_2}}  
{   M_q \over   (2\pi (  r_1 d_n - r_2 c_n   )   )^{2q_1}(2\pi (   r_1b_n-r_2 a_n   ))^{2q_2}} 
=\nn\\
&&
\nn\\
&&={2\over (b_n)^{2q_2}(d_n)^{2q_1}}\sum^{\infty}_{r_1 d_n >r_2 c_n; r_1b_n >r_2 a_n}   
 {  M_p \over   (2\pi r_1)^{2p_1}(2\pi r_2)^{2p_2}}  
  {  M_q  \over   (2\pi (r_1 - r_2 {c_n \over d_n} )^{2q_1}(2\pi (r_1  - r_2 {a_n \over b_n}))^{2q_2}} \leq \nn\\
&& \leq{1\over (b_n)^{2q_2}(d_n)^{2q_1}}\sum^{\infty}_{r_1,r_2 =1}   
 {  2 M_p \over   (2\pi r_1)^{2p_1}(2\pi r_2)^{2p_2}}  
  {  9 M_q  \over   (2\pi )^{2q_1}(2\pi)^{2q_2}} \leq \nn\\
  &&\leq e^{-n (2q_1 +2q_2) \ln \lambda } ~\sum^{\infty}_{r_1,r_2 =1}   
 {  2 M_p \over   (2\pi r_1)^{2p_1}(2\pi r_2)^{2p_2}}  
  {  9M_q  \over   (2\pi )^{2q_1}(2\pi)^{2q_2}}.\nn
 \eea
Thus for the third  term we have
\be
 {1\over 8} \vert  \sum^{\infty}_{r_1 d_n >r_2 c_n; r_1b_n >r_2 a_n} 
a_{r_1,r_2} b_{ r_1 d_n - r_2 c_n, r_1b_n-r_2 a_n} \vert ~~~ \leq~~~~ C(f,g)~e^{-n   h(T) \nu(f,g)},
\ee
where the  numerical factors are 
\bea
C(f,g) =~\sum^{\infty}_{r_1,r_2 =1}   
 {  2 M_p \over   (2\pi r_1)^{2p_1}(2\pi r_2)^{2p_2}}  
  {  9M_q  \over   (2\pi )^{2q_1}(2\pi)^{2q_2}},~~~   \nu(f,g)= 2q_1 +2q_2, \nn 
  \eea
and they are $T$  independent.  The fourth term has an identical upper bound: 
\bea
 {1\over 8} \vert  \sum^{\infty}_{ i_2 c_n>i_1 d_n; i_2 a_n >i_1b_n} 
a_{i_1,i_2}
  b_{ -i_1 d_n + i_2 c_n, -i_1b_n+i_2 a_n}\vert ~~~~
  \leq ~~~ C(f,g)~ e^{-n   h(T)  \nu(f,g)} . \nn
  \eea
We arrive to the following upper bound on the correlation function:
 \be\label{estimate}
\vert D_n(f,g) \vert \leq 
~C_1 e^{-n h(T)  \nu_1 }    
+~C_2 e^{-n  h(T)  \nu_2} 
+  ~ 2 C  e^{-n  h(T)  \nu }~.
\ee
It follows that  the dependence on the system dynamics $T$ appears in the 
exponential factor $e^{-n h(T) \nu }$ through its fundamental 
characteristic, the entropy $h(T)$.
The coefficients  $C_i(f,g), C(f,g)$ and $\nu_i(f,g), \nu(f,g)$ depend only on
observables through their smoothness indices $p_i$ and $q_i$ and 
the upper bounds on derivatives $M_p$ and $M_q$.

 {\it The above calculation provides a qualitative understanding of how 
 the  
 exponential decay of the correlation functions appears and its rate.  Under repeated action of the 
dynamic system $T$ on the observable $g(T^n x)$ its oscillating 
frequencies  are
stretching apart toward the high frequency modes and the overlapping integral 
with the fixed observable $f(x)$ falls exponentially}.   

In order to estimate the upper bound (\ref{estimate}) let us consider the observables 
of the same order of smoothness $p_1=p_2=q_1=q_2 =p$. In that case the 
formula simplifies: 
\be\label{simbound}
\vert D_n(f,g) \vert \leq 
4 C(f,g)~  e^{-  n h(T)  \nu  }  ~~,
\ee 
where 
\bea
\nu = 4p,~~~~~~4 C(f,g) ={72 M^2_p  \over   (16\pi^4 )^{2 p} }  ~\Bigg(\sum^{\infty}_{r=1}   
 {1  \over   r^{2p}}  \Bigg)^2  ~ .\nn
\eea
This result  allows to define the decorrelation time for the physical observable $f(x)$ as in \cite{krilov,yer1986a,body}:
\be
\tau_0 = {1\over  h(T) \nu_f}~~.
\ee
The 
value of the standard deviation  $\sigma_f$ is a sum
\bea
\sigma^2_f = \sum^{+ \infty}_{n=-\infty} [\langle f(T^n x) f(x)  \rangle-
\langle f(x)  \rangle^2 ]\nn
\eea
and using the (\ref{simbound})  one can 
explicitly  estimate the standard deviation $\sigma_f$  
\be
\sigma^2_f \leq  
4 C_f~ {1+e^{ -h(T) ~\nu_f } \over 1- e^{  -h(T) ~\nu_f}  }.
\ee  

\section{\it High Dimensional C-systems }

Let us now consider the operators $T$ of high dimension, $N>2$ in (\ref{eq:matrix}). 
In that case the characteristic polynomial is of high order and it is difficult 
to find out the general analytical expression for the coefficients of the 
operator $T^n$ similar to the formula (\ref{eqmatrix2}). We shall use a 
computer to calculate the coefficients of the matrix $T^n$. What we need 
is the rate at which the coefficients grow as a function  of $n$. 
The numerical experiments demonstrate that the fastest growing 
coefficients in each row are the ones which are the next to the last column and for the columns they are on the last row.  
Let us consider $N=3$:
\be
\label{eq:matrix3}
T =
   \begin{pmatrix}
      1 & 1 & 1   \\
      1 & 2 & 1   \\
            1 & 2 & 2   
   \end{pmatrix},~~~~T^n =
   \begin{pmatrix}
      a^{11}_n & a^{12}_n & a^{13}_n   \\
        a^{21}_n & a^{22}_n & a^{23}_n   \\
          a^{31}_n & a^{32}_n & a^{33}_n    
   \end{pmatrix}.
\ee
There is only one eigenvalue which is outside of the unit circle $\lambda = 4.0796..$. 
The largest coefficients in each row are $a^{12}_n, a^{22}_n,a^{32}_n$ correspondingly
and the largest coefficients in each column  are $a^{31}_n, a^{32}_n,a^{33}_n$ correspondingly, and they all grow at the rate $\propto \lambda^n$.  The observables are 
\bea
f(x) &=&  \sum^{\infty}_{i_1,i_2,i_3=1}  a_{i_1 i_2i_3} \cos(2 \pi i_1  x_1  )
\cos(2 \pi i_2   x_2)\cos(2 \pi i_3   x_3),  \\
 g(T^n x) 
&=&\sum^{\infty}_{j_1,j_2,j_3=1} b_{j_1 j_2j_3}
\cos(2 \pi j_1 ( a^{11}_n x_1+ a^{12}_n x_2 + a^{13}_n x_3 ) )
\cos(2 \pi j_2 (a^{21}_n x_1+ a^{22}_n x_2+ a^{23}_n x_3)) \nn  \\
&&~~~~~~~~~~~~~~~~~~~~~~~~~~~~~\cos(2 \pi j_3 (a^{31}_n x_1+ a^{32}_n x_2+ a^{33}_n x_3)). \nn
\eea
A typical term in the correlation function $D_n(f,g)$ will have a form  similar to the 
(\ref{correlator2}):
\be
\sum^{\infty}_{j_1,j_2,j_3=1} 
a_{j_1 a^{11}_n + j_2 a^{21}_n+j_3 a^{31}_n,~ j_1 a^{12}_n + j_2 a^{22}_n+j_3 a^{32}_n, ~
j_1 a^{13}_n + j_2 a^{23}_n+j_3 a^{33}_n}~~ b_{j_1 j_2j_3},
\ee
and it can be bound from above: 
\bea
&&
\vert  \sum^{\infty}_{j_1,j_2,j_3=1} 
a_{j_1 a^{11}_n + j_2 a^{21}_n+j_3 a^{31}_n,~ j_1 a^{12}_n + j_2 a^{22}_n+j_3 a^{32}_n, ~
j_1 a^{13}_n + j_2 a^{23}_n+j_3 a^{33}_n}~~ b_{j_1 j_2 j_3} \vert
  \leq \nn\\
&&{2\over (a^{31}_n)^{2p_1}(a^{32}_n)^{2p_2}(a^{33}_n)^{2p_3}}\sum^{\infty}_{r_1,r_2,r_3=1}  {  M_p  \over   (2\pi r_3)^{2p_1} (2\pi r_3)^{2p_2} (2\pi r_3)^{2p_3}  }{  M_q \over   (2\pi r_1)^{2q_1}(2\pi r_2)^{2q_2}(2\pi r_3)^{2q_3}}    \leq \nn\\
&& e^{-(2p_1 + 2p_2 + 2p_3) n \ln \lambda}~ \sum^{\infty}_{r_1,r_2,r_3=1}   {  M_p  \over   (2\pi r_3  )^{2p_1+2p_2 +2p_3}}  { 2  M_q \over   (2\pi r_1)^{2q_1}(2\pi r_2)^{2q_2}(2\pi r_3)^{2q_3}}   ~~,\nn
\eea
where $\ln \lambda = h(T)$ and
\be
\vert \partial^{(2p_1)}_{x_1} \partial^{(2p_2)}_{x_2}
\partial^{(2p_3)}_{x_3} f(x_1,x_2,x_3) \vert \leq M_p, ~~~~~~~
\vert \partial^{(2q_1)}_{x_1} \partial^{(2q_2)}_{x_2} 
\partial^{(2p_3)}_{x_3}g(x_1,x_2,x_3) \vert \leq M_q ~.
\ee
One can estimate the upper bound on the correlation function considering
for simplicity  the observables 
of the same order of smoothness $p_i=q_i =p$. In that case the 
formula takes the following form:
\be\label{simbound}
\vert D_n(f,g) \vert \leq 
C(f,g)~  e^{-  n h(T)  \nu  }  ~~,
\ee 
where 
\bea
\nu = 6p,~~~~~ C(f,g) ={ M^2_p  \over   (2\pi )^{12 p} }  ~\Bigg(\sum^{\infty}_{r=1}   
 {1  \over   r^{2p}}  \Bigg)^3  ~ .\nn
\eea
Generalising this calculation to the operators $T$ on dimension $N$ 
we shall get 
\bea
\nu_f = 2p N,~~~~~ C(f,g) ={ M^2_p  \over   (2\pi )^{4 p N} }  ~\Bigg(\sum^{\infty}_{r=1}   
 {1  \over   r^{2p}}  \Bigg)^N  ~ ,\nn
\eea
and the formula for the decorrelation  time takes the form 
\be
\tau_0 = {1\over  \nu_f \ h(T) }~=~ {1\over 2p N  h(T) }.
\ee
{\it The exponential decay of the correlation functions 
is getting   faster as the 
dimension $N$ of the operators $T$ is increasing.  Taking into consideration 
that the entropy $h(T)$ of our system is linearly increasing with 
$N$ ,  $h(T)\approx
{2\over \pi}~ N
$ in (\ref{linear}), we shall get the following expression 
for the decorrelation time }:
\be\label{decorrelation}
\tau_0 =   {\pi \over 4 p  N^2  }~.
\ee
Considering a set of initial trajectories occupying a small volume $\delta v_0$
in the phase space of the C-system it is important to know how fast the
volume $\delta v_0$ will be spread/distributed over the whole phase space during the evolution of the system.  This  characteristic 
time interval $\tau$ defines the time at which the system reaches a stationary distribution. 
The entropy of the system defines the expansion rate of the 
phase space volume elements and one can derive therefore that  
\cite{yer1986a}
\be\label{stationary}
\tau = {1\over h(T)} \ln {1\over \delta v_0}.
\ee  
Thus there are three characteristic time scales associated with the C-system
\cite{yer1986a}:
 \be\label{times}
  \begin{pmatrix}
 Decorrelation~ time~\\
 \tau_0 ={\pi \over 4 p N^2}
\end{pmatrix}  < 
\begin{pmatrix}
 Interaction~ time~  \\
 t_{int} =1
\end{pmatrix} <
\begin{pmatrix}
 Stationary~ distribution~ time  \\
\tau={1\over h(T)} \ln {1\over \delta v_0}
\end{pmatrix}.
 \ee
This result defines important physical characteristics of the C-systems 
and measures the "level of chaos" developed in the system and justifies 
the physical conditions at which the statistical/probabilistic description of the system is available.

\section{\it MIXMAX Random Number Generator}

One of the interesting applications of the Anosov C-systems (\ref{eq:matrix}) 
is  associated with the so called MIXMAX generator of pseudorandom numbers 
\cite{yer1986a,konstantin,Savvidy:2015ida,Savvidy:2015jva}. 
It was demonstrated in \cite{konstantin} that the MIXMAX pseudorandom
number generators are passing strong statistical U01-tests \cite{pierr} when the entropy 
of the generators is larger than fifty, $h(T) > 50$. Using the  formulas 
(\ref{decorrelation}), (\ref{stationary}) and (\ref{times})  of the last section
 one can estimate characteristic time scales associated with the MIXMAX generators.
The generator $N=256$ in Table 1 of the article \cite{Savvidy:2015jva}
has the entropy $h(T)=194$  and the smallest phase volume is of order 
$\delta v_0 = 2^{- 61 \cdot 256}$, therefore the characteristic  time scales 
for this generator are 
\be\label{times256}
  \begin{pmatrix}
 Decorrelation~ time~\\
 \tau_0 =0.000012
\end{pmatrix}  < 
\begin{pmatrix}
 Interaction~ time~  \\
 t_{int} =1
\end{pmatrix} <
\begin{pmatrix}
 Stationary~ distribution~ time  \\
\tau=95
\end{pmatrix}.
 \ee
The MIXMAX generator which has much higher entropy was presented in 
Table 3 of the article \cite{Savvidy:2015jva}. It has the entropy 
$h(T)=8679$  and the smallest volume 
$\delta v_0 = 2^{- 61 \cdot 240}$, therefore the characteristic  time scales 
for this generator are 
\be\label{times240}
  \begin{pmatrix}
 Decorrelation~ time~\\
 \tau_0 =0.000004
\end{pmatrix}  < 
\begin{pmatrix}
 Interaction~ time~  \\
 t_{int} =1
\end{pmatrix} <
\begin{pmatrix}
 Stationary~ distribution~ time  \\
\tau=1.17
\end{pmatrix}.
 \ee
Both generators have very short decorrelation time.  The second 
generator $N=240$ has much bigger entropy and therefore its 
relaxation time $\tau$ is much smaller, of order $1.17$,  
and is close to the interaction time.
In that sense it has very strong stochastic/chaotic properties, 
it much faster spreads trajectories over the whole phase space 
and reaches the equilibrium.  
Therefore it should not be surprising that these generators 
are passing all the tests in the BigCrush U01-suite \cite{pierr}.  
These generators have the best combination of speed, reasonable size of the state and are currently available  generators in the ROOT and CLHEP software packages at CERN for Monte-Carlo simulations and scientific calculation \cite{hepforge,root,clhep}.

\section{\it Conclusion}

Our analyses of the $N$ dimensional hyperbolic Anosov C-system (\ref{eq:matrix}) 
indicates that its basic statistical characteristics are expressible in terms 
of entropy. The decorrelation time $\tau_0$  and the relaxation time 
$\tau$ are inversely proportional 
to the entropy of the system and indicate that these time scales become 
shorter as entropy increases. This is an intuitively appealing result because the entropy measures the uncertainty in the description of the 
physical systems and here it is translated into the important time scales
characteristics. As a result a perfectly deterministic dynamical system 
shows up a fast thermalisation and well developed statistical 
properties.  When measuring different observables of the hyperbolic 
Anosov C-system it will be difficult to recognise that in reality the 
data are coming out from a perfectly deterministic dynamical system.  

The exponential decay of the correlation functions has been found earlier 
in classical dynamics of the N-body gravitating systems and can be 
used to justify a 
statistical description of globular clusters and elliptic galaxies 
\cite{body,Gibbons:2015qja,Chanda:2016aph}.

It was suggested in the literature that the outgoing Hawking radiation 
\cite{Hawking:1974rv} may be not exactly thermal, but had subtle correlations \cite{tHooft:1990fkf,Susskind:1993mu,Hawking:2005kf,Gur-Ari:2015rcq}.
In that respect one can suppose 
that the effective description of the black hole radiation can be understood  
in analogy with the behaviour of the hyperbolic systems of the type considered above \cite{Polchinski:2015cea}.

\section{\it Acknowledgement }
This work was supported in part by the European Union's Horizon 2020
research and innovation programme under the Marie Sk\'lodowska-Curie
Grant Agreement No 644121.

\vfill


\begin{thebibliography}{99}




\bibitem{anosov}  D. V. Anosov, \emph{Geodesic flows on closed Riemannian manifolds with negative curvature},  Trudy Mat. Inst. Steklov., Vol. {\bf 90} (1967) 3 - 210

\bibitem{kolmo} A.N. Kolmogorov,  \emph{New metrical invariant of transitive dynamical
systems and automorphisms of Lebesgue spaces},
Dokl. Acad. Nauk SSSR,  {\bf{119}} (1958) 861-865

\bibitem{kolmo1} A.N. Kolmogorov,  \emph{On the entropy per unit time as a metrical invariant
of automorphism},
Dokl. Acad. Nauk SSSR,  {\bf{124}} (1959) 754-755

\bibitem{sinai3} Ya.G. Sinai,  \emph{On the Notion of Entropy of a Dynamical System}, Doklady of Russian Academy of Sciences, {\bf{124}}  (1959)  768-771.

\bibitem{rokhlin} V.A. Rokhlin, \emph{On the endomorphisms of compact commutative groups},
Izv. Akad. Nauk, vol.  {\bf{13}} (1949) 329

\bibitem{rokhlin2}V.A. Rokhlin, \emph{On the entropy of automorphisms of compact commutative groups},
 Teor. Ver. i Pril., vol. {\bf3}, issue 3  (1961)  351


\bibitem{sinai4}Ya. G. Sinai, \emph{Proceedings of the International Congress
of Mathematicians}, Uppsala (1963) 540-559.

\bibitem{gines}A.~L.~Gines, \emph{Metrical properties of the endomorphisms on m-dimensional
torus},  Dokl. Acad. Nauk SSSR,  {\bf 138} (1961) 991-993

\bibitem{bowen0} R.Bowen, \emph{Equilibrium States and the Ergodic Theory of Anosov Diffeomorphisms}. (Lecture Notes in Mathematics, no. 470: A. Dold and B. Eckmann, editors). Springer-Verlag (Heidelberg, 1975), 108 pp.
 
\bibitem{ruelle} D. Ruelle, \emph{ Thermodynamic Formalism}, Addison-Wesley, Reading, Mass., 1978

\bibitem{krilov}N.S.Krylov, \emph{Works on the foundation of statistical physics}, M.- L. Izdatelstvo
Acad.Nauk. SSSR, 1950; (Princeton University Press, 1979)



\bibitem{yangmillsmech} G.Savvidy, \emph{The Yang-Mills mechanics as a Kolmogorov K-system}, Phys.Lett.B {\bf{130}} (1983) 303

\bibitem{yer1986a}  G. Savvidy and N. Ter-Arutyunyan-Savvidy, \emph{ On the Monte Carlo simulation of physical systems}, J.Comput.Phys. {\bf 97} (1991) 566; Preprint EFI-865-16-86-YEREVAN, Jan. 1986. 13pp.

\bibitem{konstantin} K.Savvidy, \emph{The MIXMAX random number generator},
Comput.Phys.Commun. 196 (2015) 161-165. (http://dx.doi.org/10.1016/j.cpc.2015.06.003);
 arXiv:1404.5355


\bibitem{Savvidy:2015ida}
  G.~Savvidy,
  \emph{Anosov C-systems and random number generators,}
  Theor.\ Math.\ Phys.\  {\bf 188} (2016) 1155;
  doi:10.1134/S004057791608002X
  [arXiv:1507.06348 [hep-th]].
   

 
\bibitem{Savvidy:2015jva}
  K.~Savvidy and G.~Savvidy,
 \emph{Spectrum and Entropy of C-systems. MIXMAX random number generator,}
  Chaos Solitons Fractals {\bf 91} (2016) 33
  doi:10.1016/j.chaos.2016.05.003
  [arXiv:1510.06274 [math.DS]].
   
 
\bibitem{Savvidy:1982jk}
  G.~Savvidy,
\emph{Classical and Quantum Mechanics of Nonabelian Gauge Fields,}
  Nucl.\ Phys.\ B {\bf 246} (1984) 302.

\bibitem{body} V.Gurzadyan and G.Savvidy, \emph{Collective relaxation of stellar systems},
Astron. Astrophys. {\bf 160} (1986) 203

\bibitem{Gibbons:2015qja}
  G.~W.~Gibbons,
\emph{The Jacobi-metric for timelike geodesics in static spacetimes,}
  Class.\ Quant.\ Grav.\  {\bf 33} (2016) no.2,  025004
  doi:10.1088/0264-9381/33/2/025004
  [arXiv:1508.06755 [gr-qc]].
  
\bibitem{Chanda:2016aph}
  S.~Chanda, G.~W.~Gibbons and P.~Guha,
\emph{Jacobi-Maupertuis-Eisenhart metric and geodesic flows,}
  arXiv:1612.00375 [math-ph].
 
\bibitem{Collet}  P. Collet, H. Epstein and G. Gallavotti, 
\emph{ Perturbations of Geodesic Flows on Surfaces of Constant Negative Curvature
and Their Mixing Properties},  Commun. Math. Phys. {\bf 95} (1984) 61-112 


\bibitem{moore}C. C. Moore, \emph{Exponential decay of correlation coefficients for geodesic flows}, Group representations, ergodic theory, operator algebras, and mathematical physics (Berkeley, Calif., 1984), Math. Sci. Res. Inst. Publ., vol. 6, Springer, New York, 1987, pp. 163 - 181.  

\bibitem{chernov1}N. I. Chernov,  \emph{Ergodic and Statistical Properties of Piecewise Linear Hyperbolic Automorphisms of the 2-Torus}, Journal of Statistical Physics,  {\bf 69 } (1992) 111

\bibitem{young}L. S. Young,  \emph{ Decay of Correlations for Certain Quadratic Maps}, 
Commun. Math. Phys. {\bf 146} (1992) 123-138

\bibitem{simic}
S. Simic' \emph{Lipschitz distributions and Anosov flows},
Proceedings of the
American Mathematical Society, {\bf 124} (1996) 1869




\bibitem{chernov}N. I. Chernov,  \emph{ Markov Approximations and Decay of Correlations for Anosov Flows},  Annals of Mathematics Second Series. {\bf 147} (1998) 269-324

\bibitem{dolgopyat}D. Dolgopyat,  \emph{ On Decay of Correlations in Anosov Flows},
Annals of Mathematics Second Series,  {\bf 147}  (1998)  357-390


\bibitem{chernov2} N. Chernov and L.-S. Young
\emph{Decay of Correlations
for Lorentz Gases and Hard Balls},  ,  {\bf  } (2001) 

\bibitem{tsujiit3}  V.Baladi and M. Tsujii, 
\emph{Anisotropic H\"older and Sobolev spaces for hyperbolic diffeomorphisms},
Annales de l'Institute Fourier, {\bf  57}  (2007) 127-154
(doi: 10.5802/aif.2253 )

\bibitem{tsujiit2}  M. Tsujii, 
\emph{Decay of correlations in suspension semi-flows of angle-multiplying maps},
Ergod. Theo. and Dynam. Sys.,{\bf  28}  (2008) 291-317, doi: 10.1017/S0143385707000430

\bibitem{tsujiit1}  M. Tsujii, 
\emph{Quasi-compactness of Transfer Operators for Contact Anosov Flows}
arXiv:0806.0732v3 [math.DS] 8 Apr. 2010


\bibitem{tsujiit}  M. Tsujii, 
\emph{ Contact Anosov Flows and the FBI Transform}, arXiv:1010.0396v3 [math.DS] 22 Jun. 2011.


\bibitem{faure} F. Faure and J. Sj\"ostrand \emph{Upper bound on the density of Ruelle resonances for Anosov flows}, arXiv:1003.0513v1 [math-ph] 2 Mar. 2010


\bibitem{faure1}F. Faure and J. Sj\"ostrand \emph{Semi-classical approach for Anosov diffeomorphisms and Ruelle resonances}
arXiv:0802.1780v3 [nlin.CD] 8 Sep 2008


\bibitem{mac}M. Kac, \emph{On the Distribution of Values of Sums of the Type $\sum f(2^{k t})$},
 Annals of Mathematics, {\bf 47} ( 1946) 33-49

\bibitem{leonov}V. P. Leonov, \emph{On the central limit theorem for ergodic endomorphisms
of the compact commutative groups}, Dokl. Acad. Nauk SSSR,  {\bf 124} No: 5 (1969) 980-983

\bibitem{chernov3} N. Chernov, \emph{ Limit theorems and Markov approximations for chaotic dynamical systems},Probab. Th. Rel. Fields  {\bf 101} (1995)  321-362 (doi:10.1007$/$BF01200500)

\bibitem{rozanov}Yu. A. Rozanov, \emph{A Central Limit Theorem for Additive Random Functions},
Theory of Probability and Its Applications,   {\bf 5}  (1960) 221-223
(doi: 10.1137$/$1105022)

\bibitem{rather} Ratner, M. \emph{The central limit theorem for geodesic flows on n-dimensional manifolds of negative curvature},
Israel J. Math. {\bf 16} (1973) 181 (doi:10.1007$/$BF02757869)


\bibitem{Poincare1}H.Poincar\'e, \emph{Sur un moyen d'augmenter la 
convergence des s\'eries trigonometriques}, Bulletin Astronomique {\bf 3} (1886) 521-528


\bibitem{pierr} P. L'Ecuyer and R. Simard, \emph{TestU01: A C Library for Empirical Testing of Random Number Generators},  ACM Transactions on Mathematical Software, {\bf 33} (2007) 1-40.


\bibitem{hepforge} HEPFORGE.ORG, {http://mixmax.hepforge.org};\\
\url {http://www.inp.demokritos.gr/~savvidy/mixmax.php}
 
\bibitem{root}  ROOT, Release 6.04/06 on  2015-10-13

\bibitem{clhep} GEANT/CLHEP, Release 2.3.1.1 on 2015-11-10.

\bibitem{Hawking:1974rv}
  S.~W.~Hawking,
\emph{Black hole explosions,}
  Nature {\bf 248} (1974) 30.
  doi:10.1038/248030a0
  



\bibitem{tHooft:1990fkf}
  G.~'t Hooft,
\emph{The black hole interpretation of string theory,}
  Nucl.\ Phys.\ B {\bf 335} (1990) 138.
  doi:10.1016/0550-3213(90)90174-C
 
\bibitem{Susskind:1993mu}
  L.~Susskind and L.~Thorlacius,
\emph{Gedanken experiments involving black holes,}
  Phys.\ Rev.\ D {\bf 49} (1994) 966
  doi:10.1103/PhysRevD.49.966
  [hep-th/9308100].

\bibitem{Hawking:2005kf}
  S.~W.~Hawking,
\emph{Information loss in black holes,}
  Phys.\ Rev.\ D {\bf 72} (2005) 084013
  doi:10.1103/PhysRevD.72.084013
  [hep-th/0507171].

\bibitem{Gur-Ari:2015rcq}
  G.~Gur-Ari, M.~Hanada and S.~H.~Shenker,
\emph{Chaos in Classical D0-Brane Mechanics,}
  JHEP {\bf 1602} (2016) 091
  doi:10.1007/JHEP02(2016)091
  [arXiv:1512.00019 [hep-th]].

\bibitem{Polchinski:2015cea}
  J.~Polchinski,
\emph{Chaos in the black hole S-matrix,}
  arXiv:1505.08108 [hep-th].


  
\end{thebibliography}
\end{document}